\documentclass{aa}
\usepackage{graphicx}
\def\Ha{H$_{\alpha}$}

\def\kms{~km$\cdot$s$^{-1}$}
\def\uves{U{\tiny VES}}
\begin{document}
%
   \title{Mass loss at the lowest stellar masses\thanks{Based on observations
obtained at the European Southern Observatory using the Very Large
Telescope in Cerro Paranal, Chile (observing runs 67.C-0549(B),
69.B-0126(A), 71.C-0429(C) and 71.C-0429(D)).}}

   \subtitle{}

   \author{M. Fern\'andez\inst{1,2}
          \and
          F. Comer\'on \inst{3}
          }

   \offprints{M. Fern\'andez}

   \institute{
Max-Planck-Institut f\"ur Astronomie, K\"onigstuhl 17, D-69117
Heidelberg, Germany 
\and
Instituto de Astrof\'{\i}sica de Andaluc\'{\i}a, CSIC, Camino Bajo de
Hu\'etor 50, E-18008 Granada, Spain \\
\email{matilde@mpia-hd.mpg.de}
\and 
European Southern Observatory, Karl-Schwarzschild-Strasse 2,
D-85748 Garching, Germany \\
\email{fcomeron@eso.org}
}

   \date{Received ; accepted}

   \abstract{ 

  We report the discovery of a jet in a [SII] image of Par-Lup3-4, a
remarkable M5-type pre-main sequence object in the Lupus 3 star-forming
cloud. The spectrum of this star is dominated by the emission lines
commonly interpreted as tracers of accretion and outflows. Par-Lup3-4 is
therefore at the very low-mass end of the exciting sources of
jets. High resolution spectroscopy shows that the [SII] line profile is
double-peaked, implying that the low excitation jet is seen at a
small angle 
(probably $\ga$8$^{\circ}$) with respect to the plane of the sky. The
width of the \Ha\ line suggests a dominating contribution from the
accretion columns and from the shocks on the stellar surface. Unresolved
\Ha\ emission coming from an 
object located at 4\farcs2 from Par-Lup3-4 is detected at a position
angle $\sim$30$^{\circ}$ or $\sim$210$^{\circ}$, with no counterpart
seen either in visible or infrared images.

  We also confirm previous evidence of strong mass loss from the very
low mass star LS-RCrA~1, with spectral type M6.5 or later. All its
forbidden lines are blueshifted with respect to the local standard of
rest (LSR) of the molecular cloud at a position very close to the object
and the line profile of the [OI] lines is clearly asymmetric. Thus, the
receding jet could be hidden by a disk which is not seen edge-on.

If an edge-on disk does not surround Par-Lup3-4 or LS-RCrA~1, an alternative
explanation, possibly based on the effects of mass accretion, is
required to account for their unusually low luminosities.


   }

   \titlerunning{Jets from the substellar limit}
   \maketitle
%

\section{Introduction}



 In the last decade evidence has been found that the mass accretion model
that simultaneously explains the mass accretion and mass loss features
observed in a great variety of sources, from active galaxies to low-mass
pre-main sequence stars (classical T~Tauri stars), can be also applied
at very low stellar masses and, possibly, below the substellar limit
(e.g., Fern\'andez \& Comer\'on \cite{Fernandez01.1}, Muzerolle et
al. \cite{Muzerolle03.1}, Comer\'on et al. \cite{Comeron03.1}, Natta et
al. \cite{Natta04.1}, Barrado y Navascu\'es \& Jayawardhana
\cite{Barrado04.2}). In fact, in the framework of low mass star
formation, the {\it mass accretion - mass loss} proportionality proposed
by Cabrit et al. (\cite{Cabrit90.1}) seems to hold below the substellar
limit. Natta et al. (\cite{Natta04.1}) have studied the accretion
properties of very low mass objects, more than doubling the number of
substellar objects for which the mass accretion rate, \.M$_{acc}$, is
known. They 
confirm the trend of lower \.M$_{acc}$ for lower M$_{\ast}$, although
with a large spread, possibly due to an age effect.  This trend has been
recently confirmed for the entire substellar domain, down to nearly the
deuterium-burning limit (Mohanty et al. \cite{Mohanty05.1}). Little
information can be found for mass loss rates, but the
estimation made by Comer\'on et al. (\cite{Comeron03.1}) for the very
low mass star LS-RCrA~1 falls in the expected range of values, according
to the mentioned proportionality.


 In order to confirm that the classical T~Tauri star paradigm also
applies at the very low mass regime, it is necessary to find
out whether these objects can be the exciting sources of jets or
outflows. Here we report the discovery of a jet emanating from a M5 star
(Par-Lup3-4) and we confirm previous evidence (Fern\'andez \& Comer\'on
\cite{Fernandez01.1}, Barrado y Navascu\'es et
al. \cite{Barrado04.1}) that the M6.5, or later, star LS~RCrA-1 is
the exciting source of an outflow. Both stars
were discovered in the course of \Ha\/ surveys carried out in the Lupus
3 and R CrA regions, respectively (Comer\'on et al. \cite{Comeron03.1};
Fern\'andez \& Comer\'on \cite{Fernandez01.1}). Their strong emission at
permitted lines indicates a strong mass
accretion process, while the forbidden emissions make them very good
candidates in the search for jets or outflows.
 These two stars also share an unexpected property: they are quite
underluminous (Fern\'andez \& Comer\'on \cite{Fernandez01.1}; Comer\'on
et al. \cite{Comeron03.1}). When compared to other young objects of
similar spectral types, Par-Lup3-4 and LS-RCrA~1 happen to be fainter by
almost 4~mag and 1.8~mag, respectively. These low luminosities put both
stars on the 50~Myr isochrone on the HR diagram. This age is much older
than those estimated for the other members of their young associations,
which are well below 10~Myr.
The results reported here suggest that 
neither of these objects is obscured by an edge-on disk, thus
constraining possible explanations of the underluminosity.


\section{Observations}

  All the observations reported in this paper were carried out with the
VLT (Cerro Paranal, Chile) in service mode. The narrow-band [SII] and
\Ha\ imaging observations of Par-Lup3-4 took place on the night of
May 2, 2003. They are integrations of 950~s each using the visible imaging
and low-resolution spectrograph FORS1. We have measured a seeing of
0$\farcs$6 for the [SII] image and of 0$\farcs$7 for the \Ha\ one.

  High resolution ($R$\,=\,57,000) spectroscopy of Par-Lup3-4 was
carried out over 5 nights in 2003, from July 4 to 30, using \uves, the
{\bf U}ltraviolet and {\bf V}isual {\bf E}chelle {\bf S}pectrograph
(Kaufer et al. \cite{Kaufer03.1}). Each observation consisted of two
consecutive spectra covering the range $\lambda\lambda$~3300 to
6800~\AA\ with exposure times of 1512~s.  The slit width was always
1$\farcs$2.  Since imaging and spectroscopy were scheduled in the same
period, recognition of the Par-Lup3-4 jet in the FORS1 images was
possible only after the \uves\ observations had been obtained, thus
preventing the selection of a position angle (PA) of the \uves\ slit
matching the direction of the jet. Instead, the parallactic angle was
set so as to minimize losses due to atmospheric differential refraction,
thus resulting in our observations probing a range of PAs. The log of
spectroscopic observations is presented in Table~\ref{tab:logse}. The
third column contains the seeing values measured with the DIMM
(Differential Image Motion Monitor) during the time of the exposures;
such values are stored in the ESO Observatories Ambient Conditions
Database\footnote{http://archive.eso.org/asm/ambient-server}. The range
of parallactic angles covered, as well as the average one for each pair
of exposures, are also listed.

  Narrow-band imaging observations of LS-RCrA~1 using the same setup as
for Par-Lup3-4 were carried out on June 2, 2003. High resolution
spectroscopic observations of LS-RCrA~1 were carried out from June 3 to
July 4, 2003. The instrumental setup was the same as for Par-Lup3-4.

  The spectra were reduced and analyzed using IRAF\footnote{IRAF is
distributed by the National Optical Astronomy Observatories, which are
operated by the Association of Universities for Research in Astronomy,
Inc. (AURA), under cooperative agreement with the National Science
Foundation, USA.}, paying special attention to the small spatially extended
structure present in some of the spectra. All velocities are referred to
the local standard of rest (LRS).

\begin{table*}
\caption[]{Log of the spectroscopic observations.}
\label{tab:logse}
\begin{tabular}{rccccc} \hline \hline
Date & Time & Seeing & \multicolumn{3}{c}{Parallactic angle} \\
 & Start (UT) &  & Range &  \multicolumn{2}{c}{Average}  \\ \hline
\multicolumn{1}{l}{Par-Lup3-4}   & h  \hskip 0.1cm m  \hskip 0.1cm s  &      &              &        &
 \\ \hline
$\left . \begin{array}{c}  4 \; {\rm July} \; 2003 \\ 4\;{\rm July}\;2003 \end{array} \right . $ &
$\left . \begin{array}{c} 00 \; 19 \; 41  \\ 00\;46\;21 \end{array} \right . $ &
$\left . \begin{array}{c}  1\farcs0-1\farcs7 \\ 1\farcs0-1\farcs7 \end{array}  \right . $ &
$\left . \begin{array}{c} $-$64\degr, $-$55\degr \\ $-$54\degr, $-$41\degr  \end{array} \right . $ &
$\left . \begin{array}{c} $-$60\degr \\ $-$48\degr \end{array} \right \} $ &
 -54\degr \\
$\left . \begin{array}{c}  10 \; {\rm July} \; 2003 \\ 10\;{\rm July}\;2003 \end{array} \right. $ &
$\left . \begin{array}{c} 00\; 47\; 40 \\ 01\; 13\; 56 \end{array} \right. $ &
$\left . \begin{array}{c}  1\farcs0-0\farcs6 \\ 1\farcs0-0\farcs6 \end{array}  \right. $ &
$\left . \begin{array}{c} $-$42\degr, $-$24\degr \\ $-$23\degr, $-$1\degr \end{array} \right. $ &
$\left . \begin{array}{c} $-$33\degr \\ $-$12\degr \end{array} \right\} $ &
 -22\degr \\
$\left . \begin{array}{c}  27 \; {\rm July} \; 2003 \\ 27\;{\rm July}\;2003 \end{array} \right . $ &
$\left . \begin{array}{c} 00\; 49\; 14 \\ 01\; 15\; 44 \end{array} \right . $ &
$\left . \begin{array}{c}  0\farcs7-1\farcs1 \\ 0\farcs7-1\farcs1 \end{array}  \right . $ &
$\left . \begin{array}{c} 14\degr, 34\degr \\ 35\degr, 50\degr \end{array} \right . $ &
$\left . \begin{array}{c} 24\degr \\ 42\degr \end{array} \right \} $ &
 33\degr \\
$\left . \begin{array}{c}  28 \; {\rm July} \; 2003 \\ 28\;{\rm July}\;2003 \end{array} \right . $ &
$\left . \begin{array}{c} 03\; 08\; 56 \\ 03\; 35\; 23 \end{array} \right . $ &
$\left . \begin{array}{c}  0\farcs4-0\farcs6 \\ 0\farcs4-0\farcs6 \end{array}  \right . $ &
$\left . \begin{array}{c} 79\degr, 84\degr \\ 84\degr, 88\degr \end{array} \right . $ &
$\left . \begin{array}{c} 81\degr \\ 86\degr \end{array} \right \} $ &
 84\degr \\
$\left . \begin{array}{c}  30 \; {\rm July} \; 2003 \\ 30\;{\rm July}\;2003 \end{array} \right . $ &
$\left . \begin{array}{c} 02\; 04\; 54 \\ 02\; 31\; 15 \end{array} \right . $ &
$\left . \begin{array}{c}  -- \\ -- \end{array}  \right . $ &
$\left . \begin{array}{c} 64\degr, 72\degr \\ 72\degr, 78\degr \end{array} \right . $ &
$\left . \begin{array}{c} 68\degr \\ 75\degr \end{array} \right \} $ &
 71\degr \\
\hline
\multicolumn{6}{l}{LS-RCrA~1} \\ \hline
$\left . \begin{array}{c}  3\; {\rm June} \; 2003 \\ 3\;{\rm June}\;2003 \end{array} \right . $ &
$\left . \begin{array}{c} 08\; 03\; 52 \\ 08\; 30\; 08 \end{array} \right . $ &
$\left . \begin{array}{c}  0.9\farcs-1\farcs7 \\ 0.9\farcs-1\farcs7 \end{array}  \right . $ &
$\left . \begin{array}{c} 53\degr, 64\degr \\ 64\degr, 72\degr \end{array} \right . $ &
$\left . \begin{array}{c} 58\degr \\ 68\degr \end{array} \right \} $ &
 63\degr \\
$\left . \begin{array}{c}  9\; {\rm June} \; 2003 \\ 9\;{\rm June}\;2003 \end{array} \right . $ &
$\left . \begin{array}{c} 08\; 16\; 12 \\ 08\; 42\; 32 \end{array} \right . $ &
$\left . \begin{array}{c}  0.8\farcs-1\farcs0 \\ 0.8\farcs-1\farcs0 \end{array}  \right . $ &
$\left . \begin{array}{c} 67\degr, 74\degr \\ 75\degr, 80\degr \end{array} \right . $ &
$\left . \begin{array}{c} 71\degr \\ 77\degr \end{array} \right \} $ &
 74\degr \\
$\left . \begin{array}{c}  13\; {\rm June} \; 2003 \\ 13\;{\rm June}\;2003 \end{array} \right . $ &
$\left . \begin{array}{c} 08\; 39\; 11 \\ 09\; 05\; 35 \end{array} \right . $ &
$\left . \begin{array}{c}  0.5\farcs-0\farcs8 \\ 0.5\farcs-0\farcs8 \end{array}  \right . $ &
$\left . \begin{array}{c} 78\degr, 83\degr \\ 83\degr, 87\degr \end{array} \right . $ &
$\left . \begin{array}{c} 80\degr \\ 85\degr \end{array} \right \} $ &
 82\degr \\
$\left . \begin{array}{c}  25\; {\rm June} \; 2003 \\ 25\;{\rm June}\;2003 \end{array} \right . $ &
$\left . \begin{array}{c} 05\; 59\; 34 \\ 06\; 25\; 45 \end{array} \right . $ &
$\left . \begin{array}{c}  1.1\farcs-1\farcs5 \\ 1.1\farcs-1\farcs5 \end{array}  \right . $ &
$\left . \begin{array}{c} 28\degr, 46\degr \\ 47\degr, 60\degr \end{array} \right . $ &
$\left . \begin{array}{c} 37\degr \\ 53\degr \end{array} \right \} $ &
 45\degr \\
$\left . \begin{array}{c}  4\; {\rm July} \; 2003 \\ 4\;{\rm July}\;2003 \end{array} \right . $ &
$\left . \begin{array}{c} 02\; 28\; 35 \\ 02\; 55\; 01 \end{array} \right . $ &
$\left . \begin{array}{c}  0.8\farcs-1\farcs4 \\ 0.8\farcs-1\farcs4 \end{array}  \right . $ &
$\left . \begin{array}{c} $-$79\degr, $-$74\degr \\ $-$73\degr, $-$66\degr \end{array} \right . $ &
$\left . \begin{array}{c} $-$76\degr \\ $-$70\degr \end{array} \right \} $ &
 -73\degr \\
\hline
\end{tabular}
\end{table*}
%



\section{Results}

\subsection{Par-Lup3-4}

\begin{figure*}
\centering
\includegraphics[width=5.5cm, angle=0]{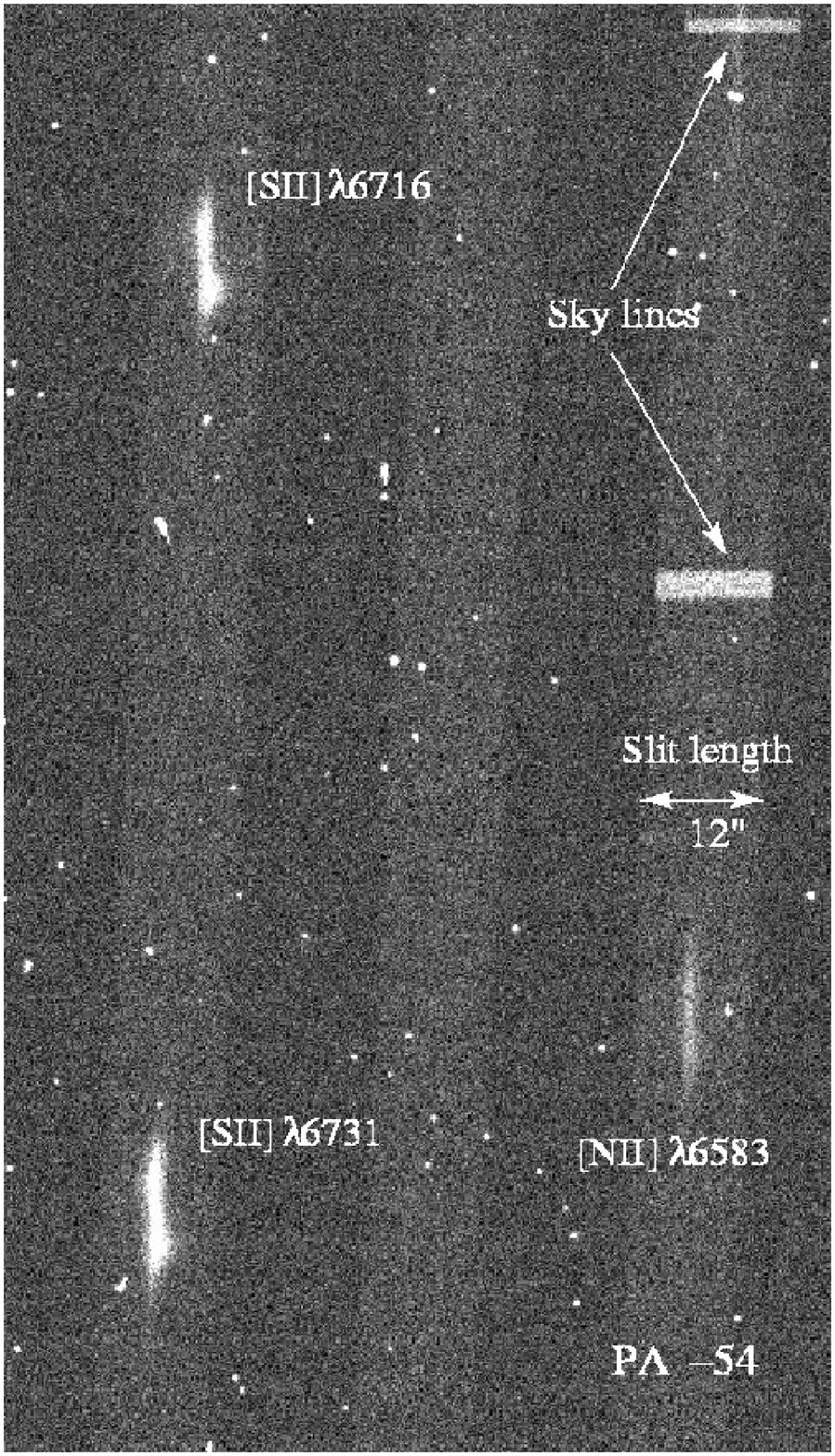}
\includegraphics[width=5.5cm, angle=0]{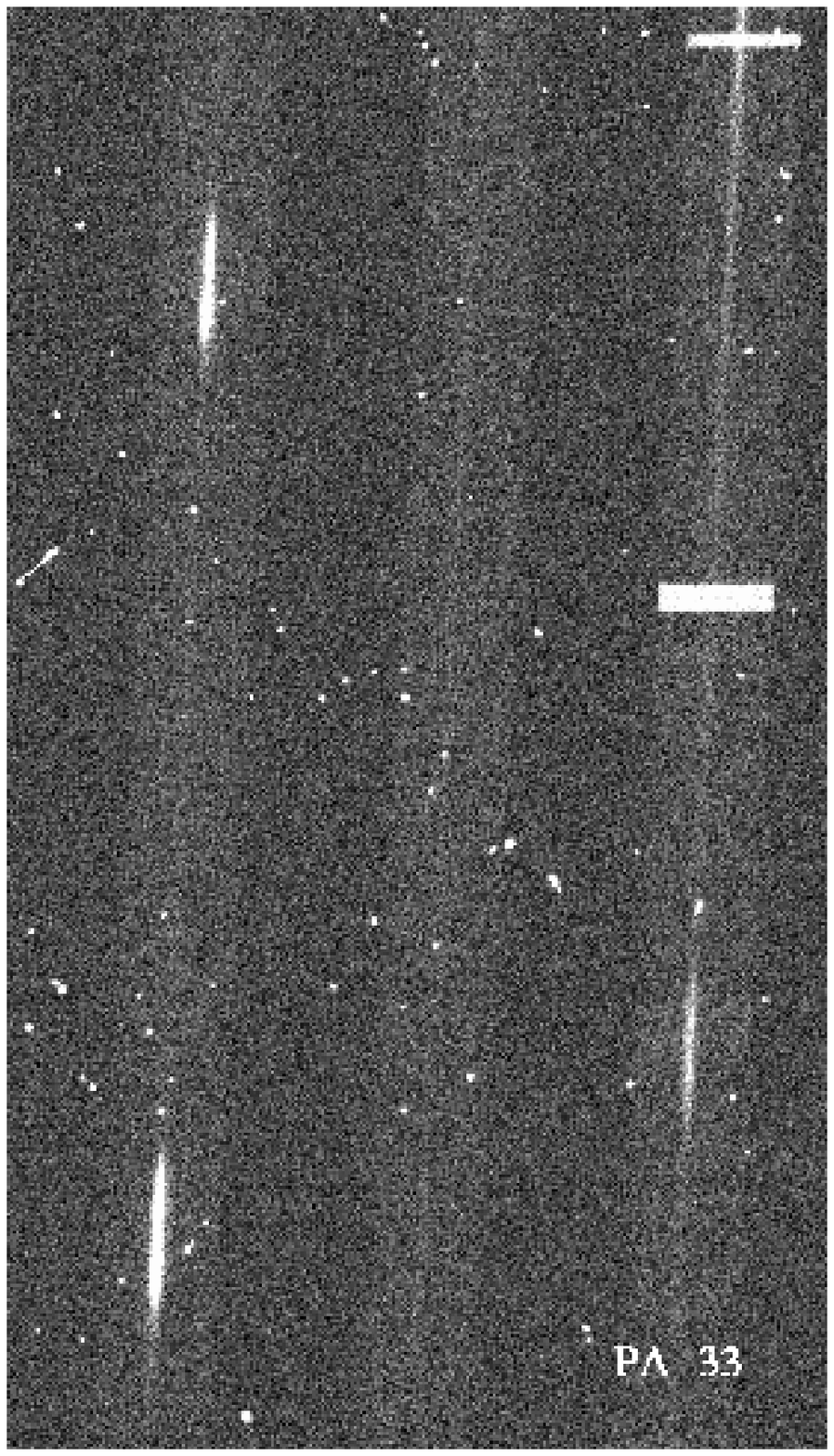}
\includegraphics[width=5.5cm, angle=0]{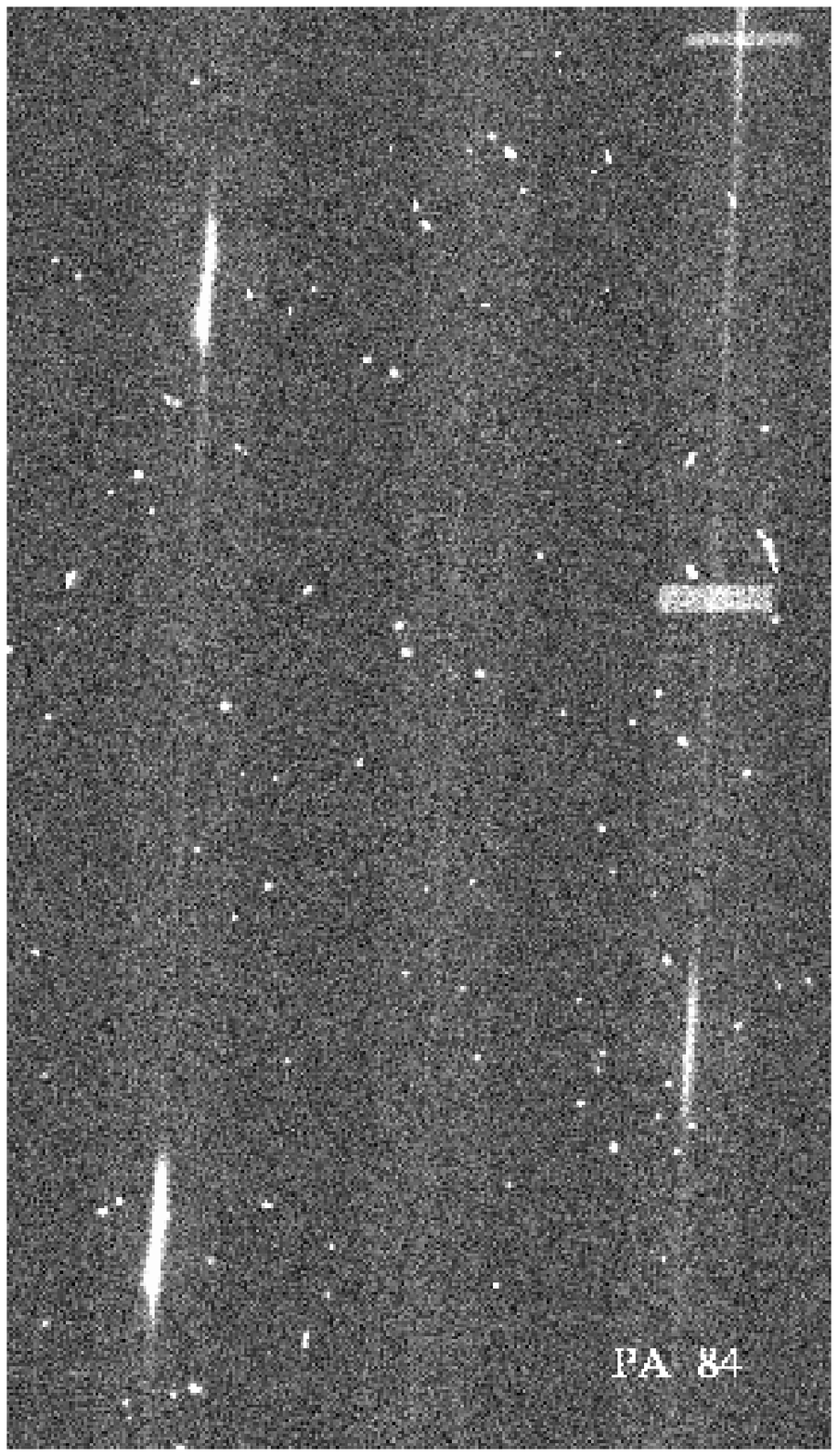}
\caption{Echellograms of Par-Lup3-4 taken on July 4 (left), July 27
  (center) and July 28, 2003 (right). Each one shows the same
sections
 of three consecutive spectral orders. The two lines on the order at the
left side of each panel are [SII] $\lambda$6716 (upper) and [SII]
$\lambda$6731 (lower); the brightest line on the order at the right side
is [NII] $\lambda$6583. 
}
\label{fig:SIIe_Par}
\end{figure*}

  Figure~\ref{fig:SIIe_Par} shows three of the echellograms obtained for
Par-Lup3-4. Each of them is the average of two consecutive
exposures. The echellogram taken on July 27 (central panel) shows 
spatially unresolved [SII] emission lines, as expected from
a point source. The data from the other two nights
show emission from the surroundings of the star. The fact that on each of
these two nights the redshifted emission (bottom of the lines)
originates at opposite sides of the star is due to the different
parallactic angles; all the emission comes, spatially, from the
same side of the star. The observed emission 
can be traced up to 3\farcs6 from the star. Hints of very
faint blueshifted emission coming from the other side of the star can be
seen on the echellogram from July 4.

\begin{figure*}
\centering
\includegraphics[width=8cm, angle=0]{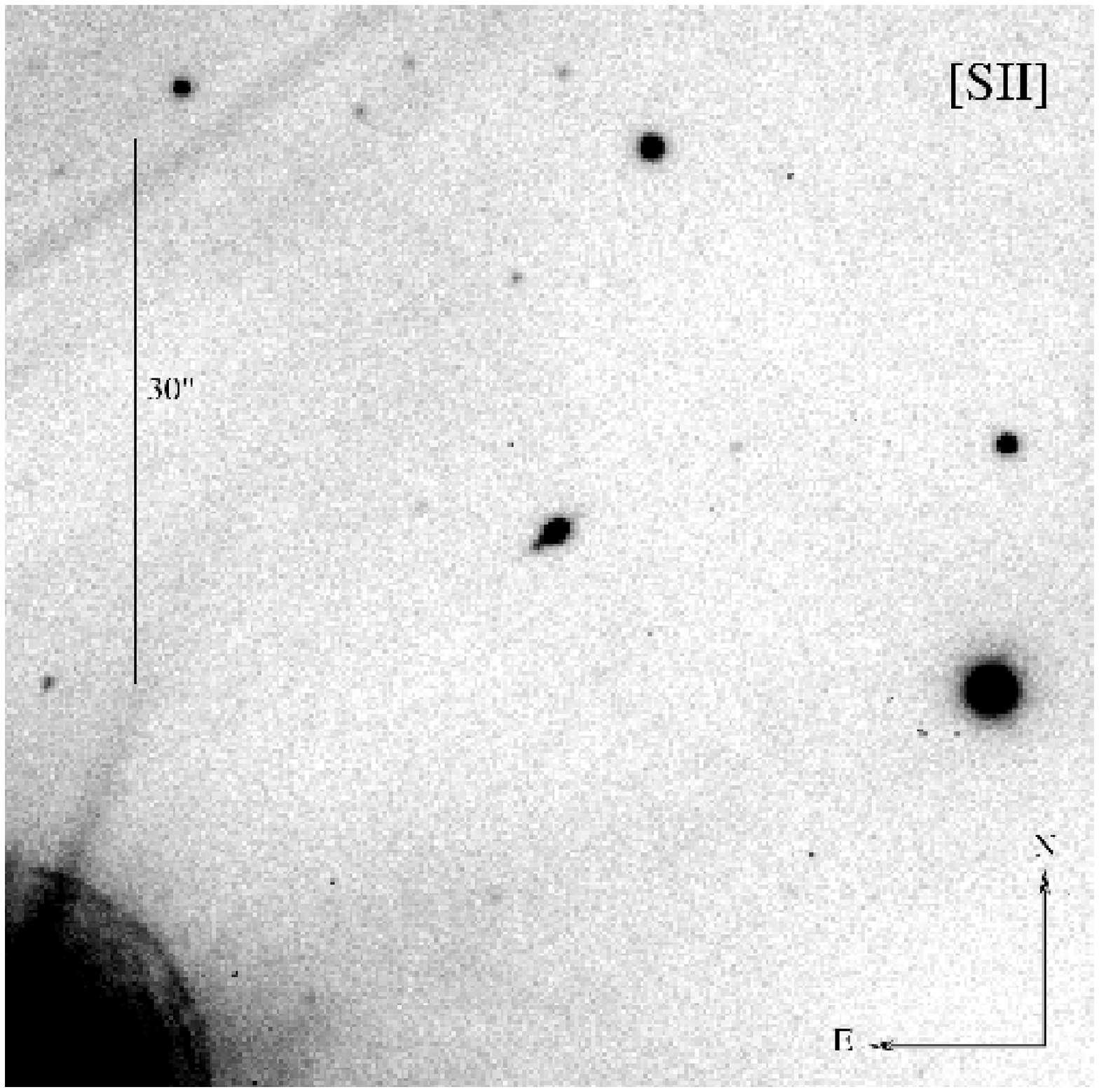}
\includegraphics[width=8cm, angle=0]{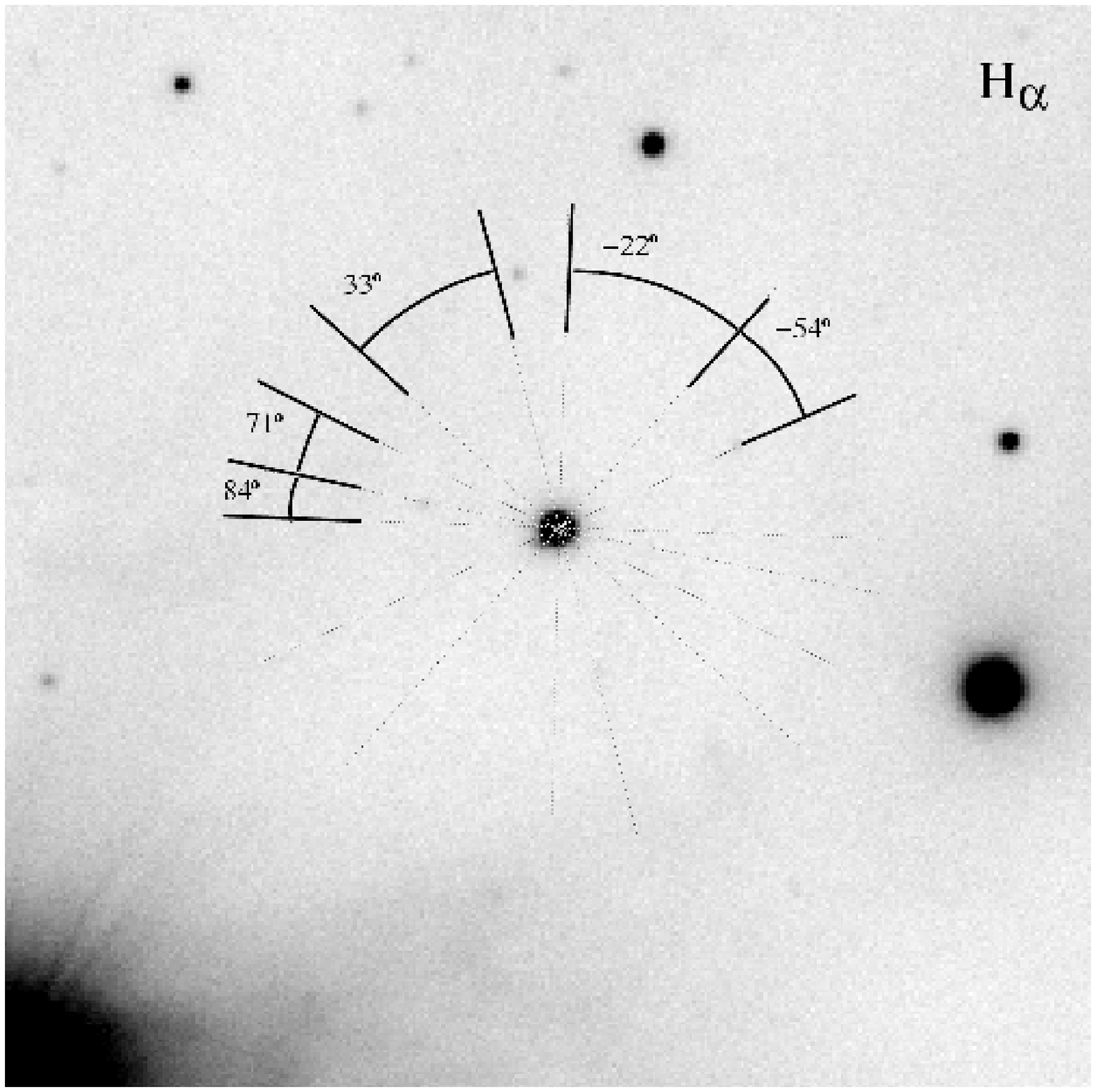}
\caption{[SII] (left) and \Ha\/ (right)
images centered on Par-Lup3-4. 
The slit position angles covered by the five pairs of spectra taken on
Par-Lup3-4 during July 2003 are plotted on the \Ha\/ image. The values of
the average slit position angles during the echelle observations are
indicated. 
}
\label{fig:SIIi_Par}
\end{figure*}

  The extended emission is clearly detected on the [SII] narrow band
image of Par-Lup3-4 (see Fig.~\ref{fig:SIIi_Par}), where a distinct knot
is seen at a PA = 129\fdg7 and at a distance of 1\farcs3 from the star,
corresponding to a 260~AU projected distance from the star assuming a
distance of 200~pc for the Lupus region; see discussion in Comer\'on et
al. (\cite{Comeron03.1}). Fainter jet-like emission can be traced
further away up to 4\farcs2 from the star (840~AU), as well as in the
opposite direction reaching up to 2\farcs0 (400~AU). The orientation of
these features implies that the slit was oriented along the jet in our
spectra of July 4, as seen on the right panel of
Fig.~\ref{fig:SIIi_Par}, where the range of slit PAs covered by the
echelle observations are plotted, giving the average PAs as a
reference. The extended emission is, nevertheless, difficult to see
on the narrow band \Ha\/ image, most probably due to the dominance of
the \Ha\ emission from the star. This extended emission is not detected
on the \Ha\/ echellogram taken along the orientation of the jet.

\begin{figure}
\centering
\includegraphics[width=4.7cm, height= 6cm, angle=0]{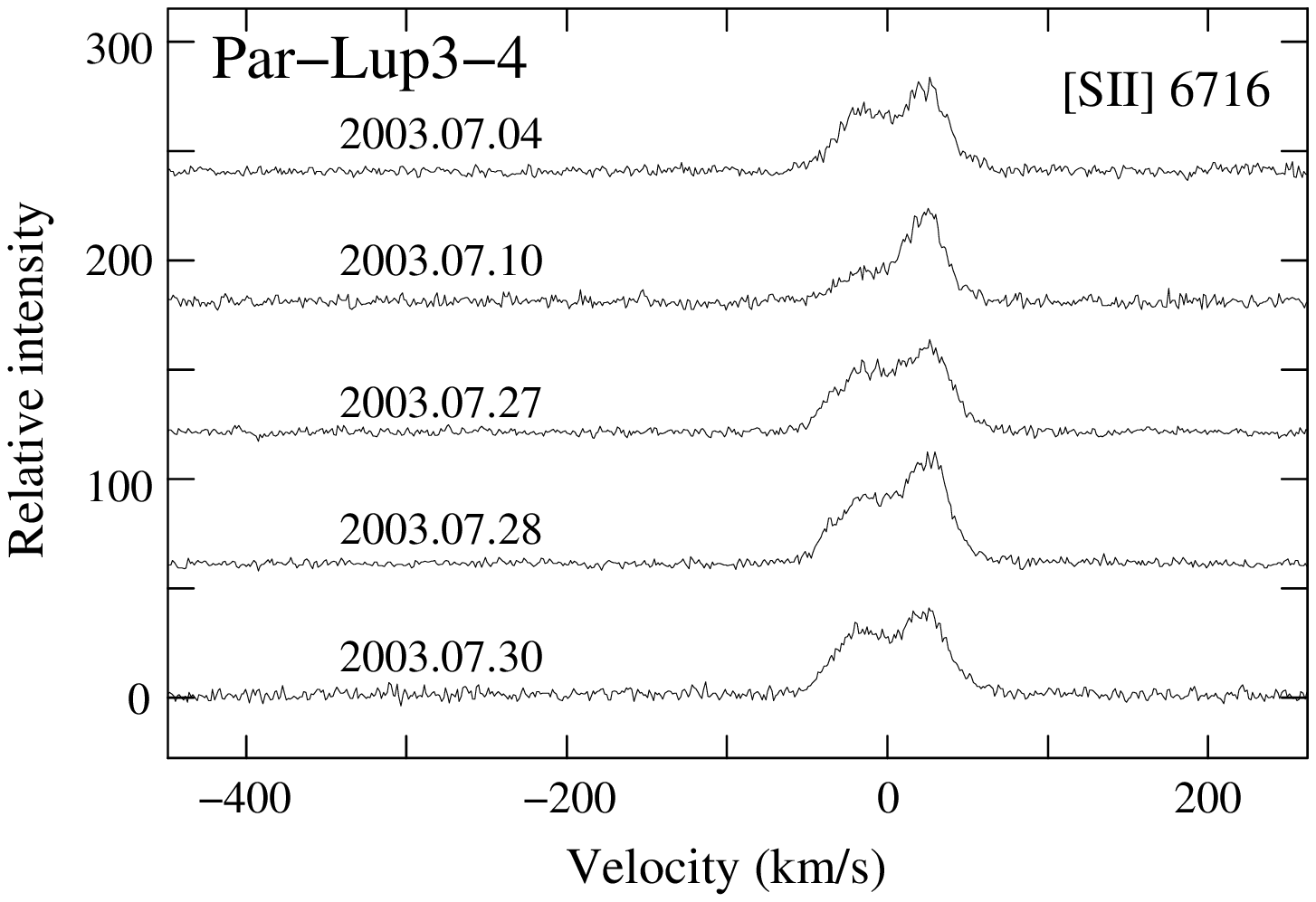}
\includegraphics[bb=100 0 468 312, width=2.5cm, height= 4.5cm,
angle=0]{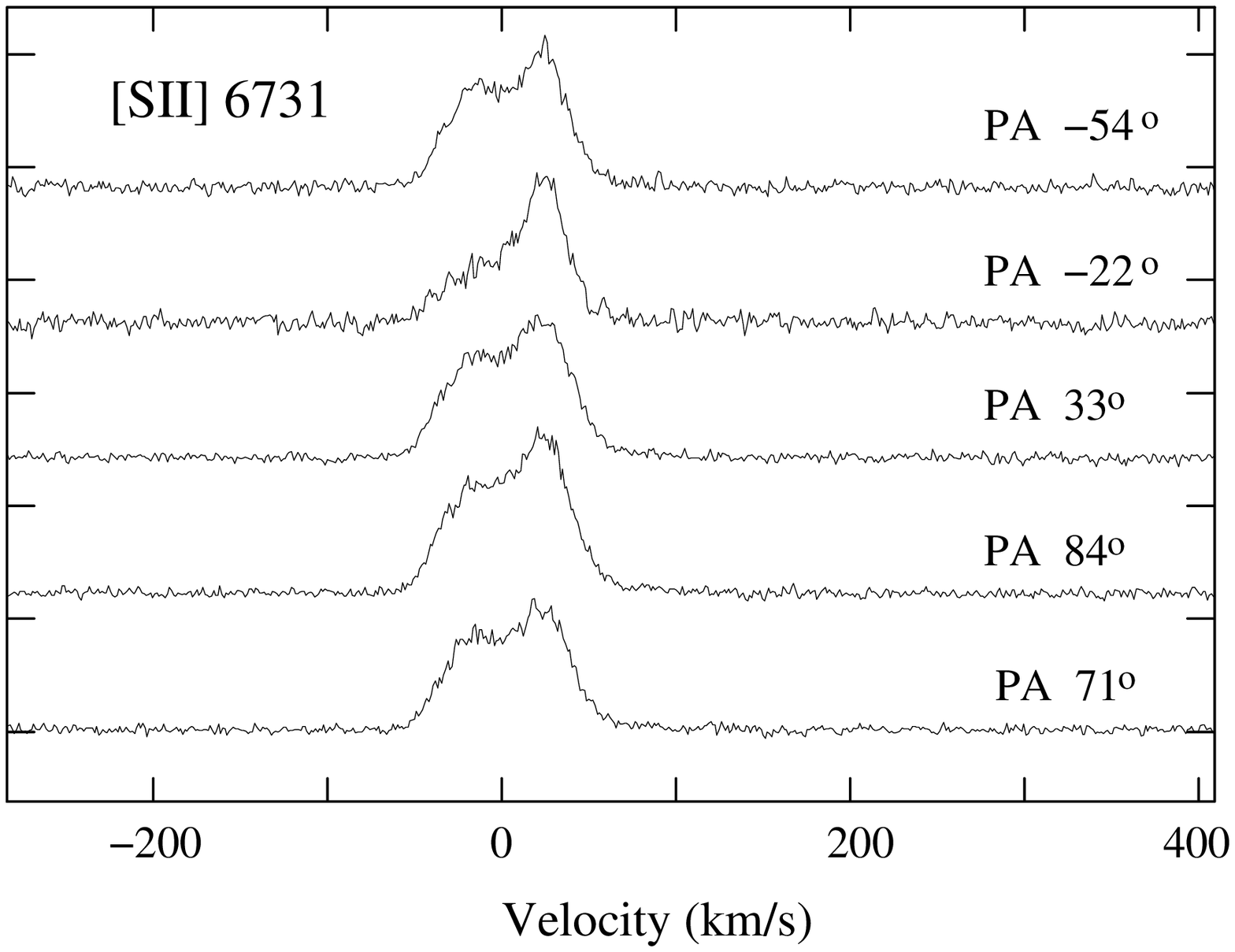} 
\caption{[SII] spectra of Par-Lup3-4. Each spectrum shows the average
of the two exposures obtained each night.} 
\label{fig:SIIech_Par}
\end{figure}

  The [SII] lines, presented in Fig.~\ref{fig:SIIech_Par}, show a
double-peaked profile, in which the relative intensities of both peaks
change. Since variability with time scales of weeks is not expected at
such a distance from the exciting source, changes are likely to be due to
the different PA of the slit probing different parts of the extended
emission. The brighter and redshifted component is associated with the
emission knot seen in the [SII] images to the south-east of Par-Lup3-4.

\begin{table*}
\caption[]{LSR velocities (km$\cdot$s$^{-1}$) of the forbidden
emissions of Par-Lup3-4 and LS-RCrA~1.}
\label{tab:veloc_Par_LS}
\begin{tabular}{lcccccc} \hline \hline
Target & [OI] $\lambda$ 6300 & [OI] $\lambda$ 6363 & [NII] $\lambda$ 6548
& [NII] $\lambda$ 6583 & [SII] $\lambda$ 6716 & [SII] $\lambda$ 6731 \\
\hline 
Par-Lup3-4 & 0.76$\pm$1.36 &  2.50$\pm$1.41 & 6.41$\pm$4.5 &  1.37$\pm$5.0 &
 -15.6$\pm$1.9, 25.0$\pm$1.5 & -18.3$\pm$1.6, 23.6$\pm$1.4 \\
LS-RCrA~1  &  -4.9$\pm$0.8 & -5.5$\pm$1.0 &  -18.3$\pm$4.37 &
-22.3$\pm$1.46 & -11.6 $\pm$ 1.7 & -13.4 $\pm$ 1.1 \\ 
\hline 
\end{tabular}
\end{table*}

  The [NII] lines are in emission in some of the spectra, but they are
absent in others. We have found no correlation between the intensity of
these lines and the slit PA, but the intensity seems to correlate with
the seeing at the time of the observations: the better the seeing, the
more intense the lines are. The spectra with better signal to noise
show a double-peaked profile. This is a clear signature of a
bipolar jet, since [NII] emission only originates in the high-velocity
component of the bipolar outflows (Hirth et al. \cite{Hirth97.1}).
The [OI] $\lambda$ 6300 and the [OI]
$\lambda$ 6363 emission lines, on the contrary, present single peak
profiles. 

 The LSR velocities measured for the [OI], [NII], and [SII] lines are
listed in Table~\ref{tab:veloc_Par_LS}. These velocities correspond to
the central position of each line or line component, in the case of
double-peaked profiles.

  We have also detected emission from the HeI~$\lambda$5876 permitted
line. Since this line is too noisy in the individual spectra, we have
added up all of them and we have measured an equivalent width of
3.7$_{-1.7}^{+0.8}$~\AA.  An equivalent width of 1.6~\AA\/ was already
reported for the HeI~$\lambda$6678 emission of this star by Comer\'on et
al. (\cite{Comeron03.1}).

\begin{figure}
\centering
\includegraphics[bb=130 0 300 312, width=3.5cm,
angle=0]{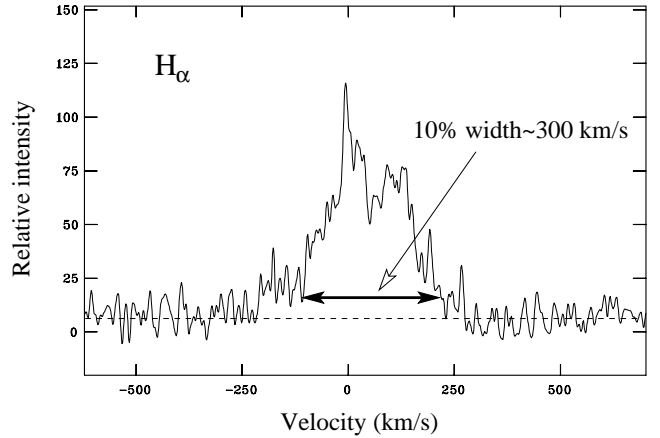}
\caption{\Ha\/ spectrum of the faint companion of Par-Lup3-4.}
\label{fig:cPar_Ha}
\end{figure}

  \Ha\/ emission, coming from an object located at 4\farcs2 from
Par-Lup3-4, has been detected in the two spectra taken at PA
24$^{\circ}$ and 42$^{\circ}$ (average PA of 33$^{\circ}$). 
Several reasons let us discard the possibility that it is an 
artifact, i.e. light reflection: the little
resemblance to the \Ha\/ profile of Par-Lup3-4; the lack of any bright
feature on the spectrum (the brightest one amounts to less than 1000
counts); the fact that its position does not change from one spectrum
to the next one, but it gets fainter, as if the object 
slowly moves out of the slit; and the fact that the \Ha\ profile of the
closest bright star (RX J1608.9 $-$3905, located at 1') is not in
emission, but filled.
From the position of this spectrum on
the echellogram we estimate a PA of $\sim$30$^{\circ}$ or
$\sim$210$^{\circ}$ for the unknown object\footnote{The uncertainty in
the PA is
due to the fact that the spectrograph is allowed to turn 360$^{\circ}$.}. 
The line that connects Par-Lup3-4 and this object subtends,
thus, an angle of $\sim$80$^{\circ}$ with the jet. An outflow direction
perpendicular to a binary axis has been reported for V536~Aql
(Mundt \& Eisl\"offel, \cite{Mundt98.1}); the distance between both
components being 0\farcs52 
($\sim$ 120 AU).  The \Ha\/ emission of the new object, presented in
Fig.~\ref{fig:cPar_Ha}, shows a wide, double-peaked profile with extended
wings covering more than 300~km$\cdot$s$^{-1}$. The two peaks are
centered at $\sim$-10~km$\cdot$s$^{-1}$ and $\sim$90~km$\cdot$s$^{-1}$,
the red peak being the faintest one. 
 No visible counterpart has been found on R, I, and z band images
up to limiting magnitudes of R=24~mag, I=22~mag, and z=21~mag (Comer\'on
et al. in preparation). 
No infrared counterpart has been found brighter than J=20.8~mag,
H=17.5~mag and 
K$_S\sim$20.8~mag in images obtained at the New Technology Telescope
(NTT), at La Silla, with SOFI in June 2001 (J band) and July 2002 (H
band), and with ISAAC at the VLT (H$_2$ filter) in August 2003.
 The line profile, as well as the
velocities of the two intensity peaks, resembles the \Ha\/ emission of
pre-main sequence objects (Fern\'andez et al. \cite{Fernandez95.1};
Reipurth et al. \cite{Reipurth96.1}). If this is the case, and taking
into account the fact that the object is more than $\sim$8~mag fainter
than the young M5 stars of this star forming region, its position on
the luminosity vs. age diagram of Burrows et al. (\cite{Burrows97.1},
see their Fig.7) would fall on the regime of the young,  
very low mass brown dwarfs and planetary mass objects. Its \Ha\/
emission could be due to mass accretion or to a flare, like the one that
has been recently observed with \uves\ on the old M9 dwarf DENIS
104814.7-395606.1 (Fuhrmeister \& Schmitt \cite{Fuhrmeister04.1}). The
flare option is, nevertheless, less plausible, because the \Ha\ intensity
does not change much from the first spectrum to the second and both
were taken over a time interval of 50~minutes. Mass accretion, on the
other hand, is known to play a very important role in the
formation of 
brown dwarfs and, possibly, of Jovian planets (Quillen \& Trilling
\cite{Quillen98.1}). A conservative value of the full width at 10\% of
the \Ha\ peak profile is in the range of 250$-$300~km$\cdot$s$^{-1}$,
which supports the hypothesis of mass accretion (see Sec. 4.1).

 Deep H$_2$ (2.12~$\mu$m) images of Par-Lup3-4 taken with ISAAC in
August 2003 show a point source located at 1$\farcs$2 from it, at a
PA$\sim$63\degr. If it has no emission lines in this band, the source is
about 6~mag fainter than Par-Lup3-4 in K$_S$, reaching
K$_S\sim$19.6~mag. If physically related to Par-Lup3-4 the object could
be in the planetary mass regime (Burrows et al. \cite{Burrows97.1}). On
the echellograms obtained on July 4, 2003, we are not able to detect
\Ha\ emission at this position, perhaps because it is dominated by the
emission from Par-Lup3-4, due to a seeing $\geq$1\arcsec.



%

%

%

\subsection{LS-RCrA~1}

\begin{figure}
\centering
\includegraphics[width=4.5cm, height= 6cm, angle=0]{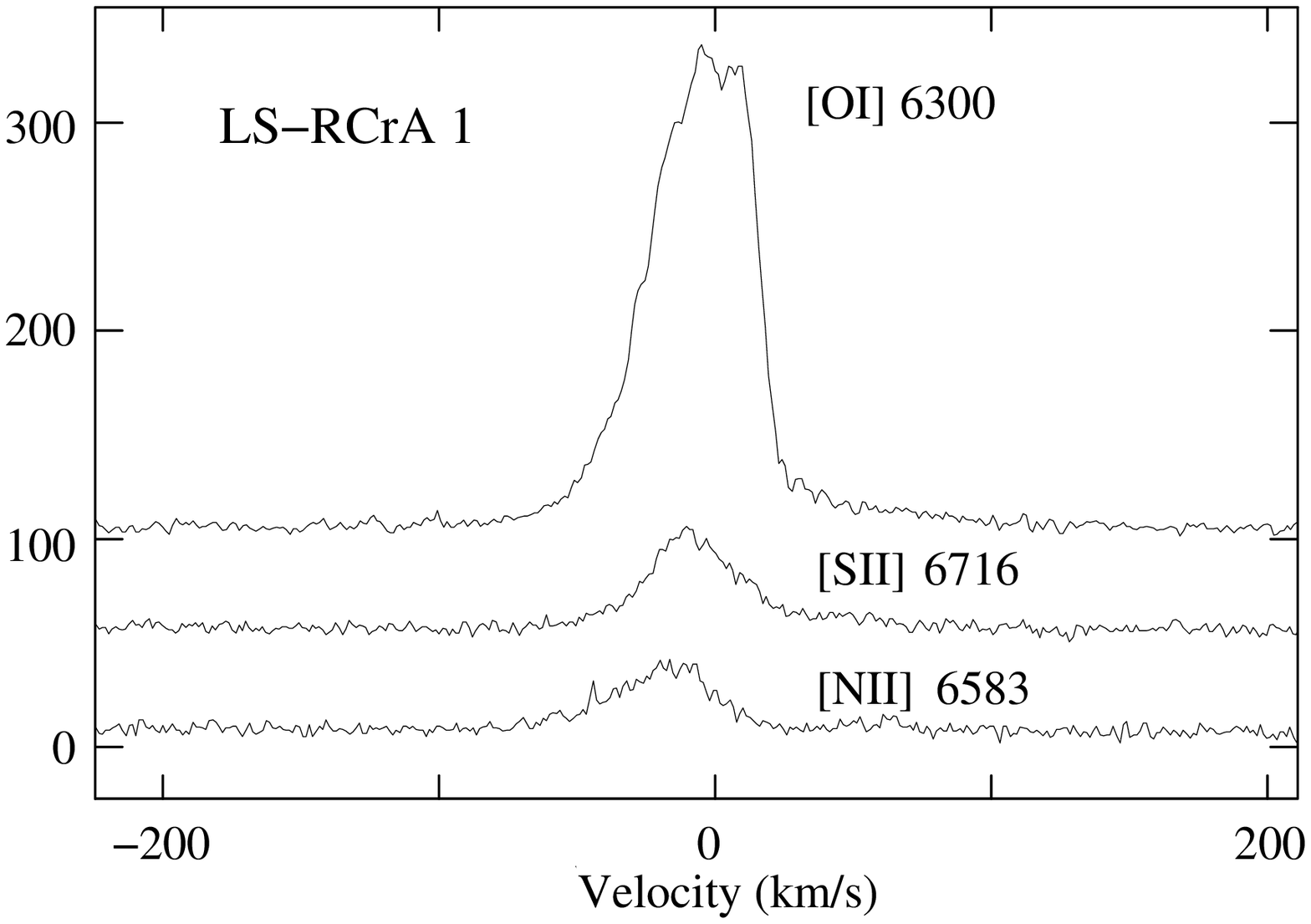}
\includegraphics[width=3.9cm, height= 6cm,
angle=0]{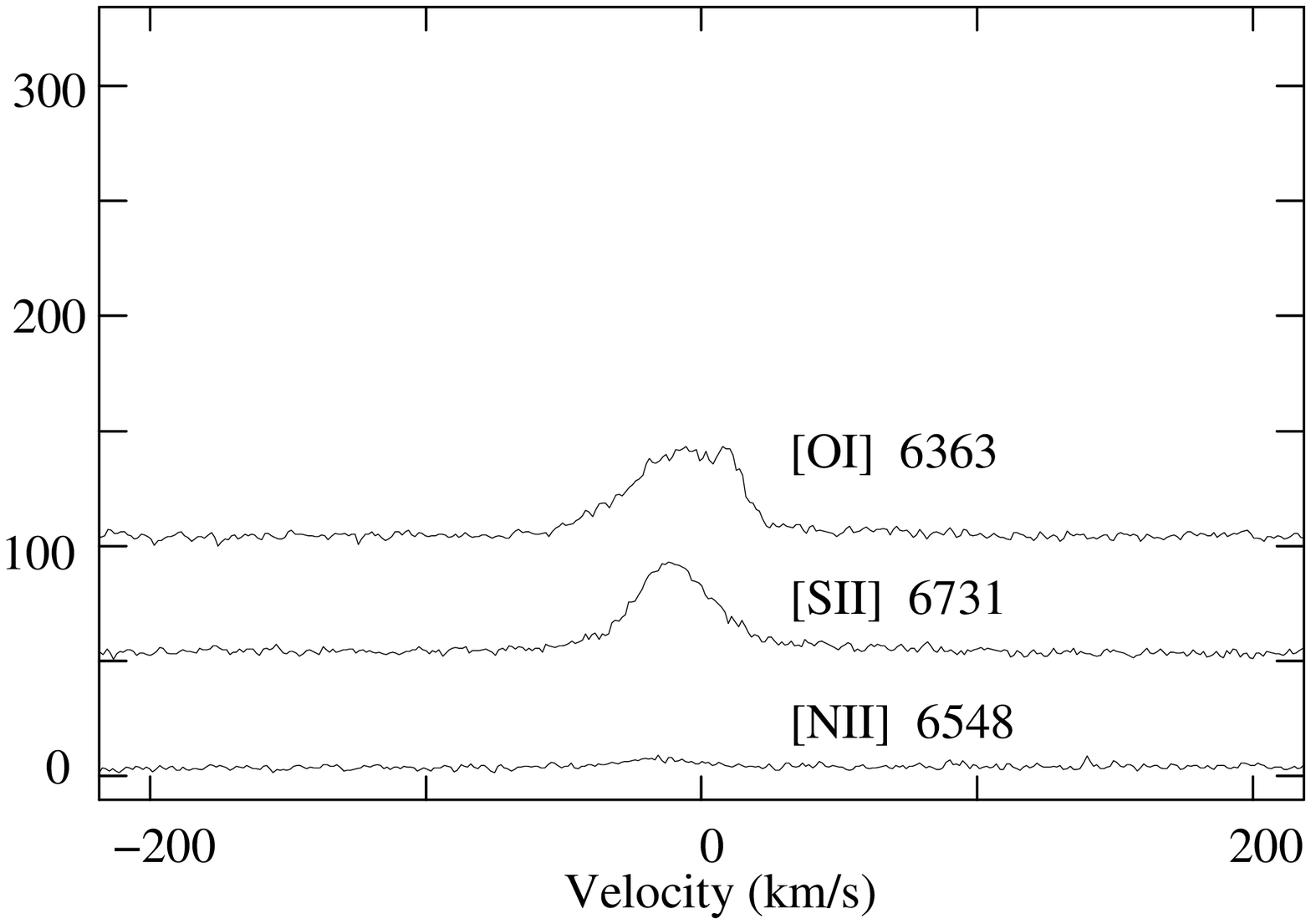}
\caption{Average forbidden line spectra of LS-RCrA~1. The central
velocity of the lines are shifted to shorter wavelengths from [OI] to [NII].
The continuum emission lies at the normalized value of 1; spectra have
been shifted for clarity.}
\label{fig:SIIe_LS}
\end{figure}

  The echellograms show no hint of extended emission close to the star
and therefore there is no dependence on the PA of the
slit. Fig.~\ref{fig:SIIe_LS} shows, for this reason, the average
forbidden lines obtained for LS-RCrA~1.

 Unlike in the case of Par-Lup3-4, no double-peaked profile is
identified in any of the forbidden lines. However, asymmetric profiles
are clearly seen. We like to note that such types of profiles are
more common among classical T Tauri stars than the double-peaked profiles
seen in Par-Lup3-4 (see Hirth et al. \cite{Hirth97.1}).
The asymmetry of the [OI] lines seems to be due to the
absorption of the redshifted emission; while the [SII] lines show a bump
on the red wing, centered at $\sim$~50~km~s$^{-1}$. The [NII] lines
present quite symmetric profiles, but both of them show faint emission
at $\sim$~50~km~s$^{-1}$. All the forbidden lines have a FWHM of about
40~km$\cdot$s$^{-1}$.

  We detect emission from both
[NII]\,$\lambda$6583 and [NII]\,$\lambda$6548, two lines that have been
reported to form only a high velocity component (HVC) and not a low
velocity component (LVC) in classical T~Tauri stars (Hirth et
al. \cite{Hirth97.1}).


%



\section{Discussion}

\subsection{Par-Lup3-4}

 As expected from its low resolution spectrum (Comer\'on et
al.~\cite{Comeron03.1}), the observed forbidden lines of Par-Lup3-4 are
characteristic of the shocked gas usually observed in jets from pre-main
sequence stars. The precise physical characteristics of Par-Lup3-4 are
difficult to determine due to its anomalously low luminosity, which
prevents its comparison with theoretical evolutionary tracks in the
temperature-luminosity diagram. Nevertheless, the late-type spectrum of
the central object places Par-Lup3-4 among the least massive
objects known to excite a jet. Other very low mass objects have been
also reported to power outflows.  Froebrich et
al. (\cite{Froebrich03.1}) found VLA 1623 to be the lowest mass star
among a sample of Class 0 sources powering outflows; they estimate that
it will reach a mass of 0.07~M$_{\odot}$. Very low masses are also
expected for BKLT J162658-241836 and WLY 2-36, the likely exciting
sources of Herbig-Haro objects in the $\rho$~Ophiuchi embedded cluster
(G\'omez et al. \cite{Gomez03.1}).


 The fact that only one side of the jet is clearly detected on the visible
images is not strange among young stars. {\it One-sided jets}, with a very
faint counterjet, have been reported for several of them (e.g.,
DG~Tau, Solf \& B\"ohm \cite{Solf93.1}). Hirth et al. (\cite{Hirth94.1})
discuss asymmetries in bipolar jets from young stars, which can be
related to the source itself or to its immediate environment.


 The lack of photospheric features on the weak continuum of the high
resolution spectra prevents us from computing the LSR velocity of the
star. Nevertheless, an estimation of this velocity 
can be obtained from the CO observations carried out by Gahm et
al. (\cite{Gahm93.1}) towards the Lupus 3 cloud, which gave an average
value between 5 and 6 km$\cdot$s$^{-1}$. More recently, Hara et
al. (\cite{Hara99.1}) measured the LSR velocities of a C$^{18}$O
core located at less than 1' from the position of Par-Lup3-4;
the beam size of the telescope was 2$\farcm$6. They measured a velocity of
4.13~$\pm$1.2~km$\cdot$s$^{-1}$. Since C$^{18}$O traces only the dense
parts of the clouds, unlike $^{12}$CO, the C$^{18}$O velocity relates
more to young stars formed very recently. The average LSR velocities
that we measure for the jet of Par-Lup3-4, 4.7$\pm$1.7~km$\cdot$s$^{-1}$
for [SII] $\lambda$6716 and 2.7$\pm$1.5~km$\cdot$s$^{-1}$ for [SII]
$\lambda$6731, match the values obtained from radio
observations, thus confirming that the red- and blue-shifted peaks
come from the jet and the counterjet, respectively.

  Gaussian fits to the double-peaked [SII] lines show an average
difference between peaks of $\simeq 41$~km$\cdot$s$^{-1}$. If we assume
a jet velocity of 150~km$\cdot$s$^{-1}$ perpendicular to the plane of
the disk, we get a disk tilt of 8\degr\, with respect to the plane of the
sky, if there is a symmetric distribution of velocities in the jet. For a
velocity of about 100 km$\cdot$s$^{-1}$ (the measured velocity width at
the base of the individual line components of the [SII] emissions) the
corresponding inclination is 12\degr. Flared disks, for which the ratio
of the disk scale height H to the radial distance R increases with R
(see Hartmann \cite{Hartmann98.1}), can hide the star more easily than
flat disks. 
To date only small samples are available for the study
of the frequency of flared disks among 
brown dwarfs; nevertheless, this frequency does not seem to be
high. From the study of the disks around 12 
brown dwarfs, Natta et al. (\cite{Natta02.1}) concluded that nine of
them might have flat disks, in spite of the strong bias of their sample
against objects with flat disks. Mohanty et al. (\cite{Mohanty04.1}),
on the other side, found strong evidence of flared disks for two brown
dwarfs out of a sample of three and Sterzik et
al. (\cite{Sterzik04.1}) reported a flare disk geometry for Cha\Ha 1. 

 The \Ha\ emission of three T~Tauri stars with edge-on disks, where the
contributions arising from the surface and its closest vicinity are
blocked from direct view, has been recently studied by Appenzeller et
al. (\cite{Appenzeller05.1}). In all cases the line has a narrow
profile, with a full width half maximum (FWHM) below
100~km$\cdot$s$^{-1}$. Appenzeller et al. interpret it as the \Ha\
contribution to the outflows.  This result strongly supports the
conclusion of White \& Basri (\cite{White03.1}), confirmed by Natta et
al. (\cite{Natta04.1}), that \Ha\ emission can only be undoubtedly
attributed to the mass accretion process, if the full width of the
emission profile at 10\% of the maximum intensity (hereafter 10\% width)
is above\footnote{For a Gaussian function the 10\% width is
1.8226 times the FWHM. Note, however, that stellar line profiles are not
always  
Gaussian.} 270~km$\cdot$s$^{-1}$. Jayawardhana et
al. (\cite{Jayawardhana03.1}) suggest, nevertheless, that for some
accreting objects the 10\% width could be as low as
200~km$\cdot$s$^{-1}$.
The wide and complex \Ha\ profile that we have observed for
Par-Lup3-4, with a 10\% width in the range from 340 to
400~km$\cdot$s$^{-1}$, should come, then, from the accretion related
regions, which lie very close to or on the stellar surface, strongly
supporting the non-edge disk hypothesis.  We also detect in our spectra
HeI~$\lambda$5876 emission, which is usually interpreted as being formed
very close to the stellar surface. However, Appenzeller et al.
(\cite{Appenzeller05.1}) have detected it also in the spectrum of their
sample of stars with edge-on disks, thus suggesting that HeI emission
can also be produced far from the surface.

 The flux ratios of the observed forbidden lines inform about the
physical characteristics of the jet. Bacciotti \& Eisl\"offel
(\cite{Bacciotti99.1}) have developed a technique which allows one to
determine the local ionization fraction, the electron density and
electron temperature using these ratios. The low signal to noise along
the jet prevents us from carrying a detailed spatial study of these line
ratios, but we can get average values for the whole jet. The observed
line ratios are

 \[ \frac{[SII] \lambda 6716}{[SII] \lambda 6731}=0.64 \pm 0.04 \left\{
\begin{array}{l}
 \mbox{blue component } 0.74 \pm 0.11 \\
 \mbox{red component }  0.58 \pm 0.06
\end{array}
\right.  \]
 \[ \frac{[SII] (\lambda 6716 + \lambda 6731)}{[OI] (\lambda 6300 +
     \lambda 6363)}=0.37 \pm 0.04  \] 
 \[ \frac{[OI] (\lambda 6300 +  \lambda 6363)}{[NII] (\lambda
     6548+\lambda 6583)} \geq 14.8 \]

\noindent From comparison to other jets for which
these lines have been measured (Bacciotti \& Eisl\"offel 1999), we can
conclude that we deal with a low excitation jet.

  The widths of the [OI] lines are very similar to those of the [SII]
lines, in contrast to what has been observed for some classical T Tauri
stars (Hirth et al. \cite{Hirth94.1}). Hirth et al. showed that the [OI]
lines form at smaller distances from the star than the [SII] lines.
The fact that the velocities involved in the formation of both sets of
lines seem to be similar could suggest that the collimation mechanism is
already working very close to the star.


\subsection{LS-RCrA~1}

 Like for Par-Lup3-4, the
non-detection of the photospheric continuum in the high 
resolution spectra 
of LS-RCrA~1 prevents us from computing the LSR
velocity of the star. Barrado y Navascu\'es et al. (\cite{Barrado04.1})
have measured a radial velocity of 2$\pm3$~km$\cdot$s$^{-1}$. This
velocity falls slightly outside of the range of velocities that
Neuh\"auser et al. (\cite{Neuhaeuser00.1}) have measured for 12 earlier
type T Tauri stars in the CrA complex (0 to -5 km$\cdot$s$^{-1}$), but
it contains within its error bars the LSR H$_2$CO velocity reported by
Loren (\cite{Loren79.1}) at a position located at less than 40" from the
star and measured with a beam size of 2$\farcm$3; for this position Loren
reported a velocity of 6.1$\pm$0.6~km$\cdot$s$^{-1}$. More recently,
Vilas-Boas et al. (\cite{VilasBoas00.1}) observed C$^{18}$O emission
from a condensation located at less than 4' from LS-RCrA~1, using a beam
size of about 1$\farcm$5, and they obtained a LSR velocity of
5.65$\pm$1.18~km$\cdot$s$^{-1}$. 

 Taking either 6 or 2~km$\cdot$s$^{-1}$ as the stellar LSR velocity, all
the measured forbidden lines are blueshifted with respect to the star
(see Table~\ref{tab:veloc_Par_LS}). This may be interpreted
as meaning that only the blueshifted component of the outflow is seen
from our vantage point, with the redshifted one probably occulted by a
circumstellar disk, as observed in most classical T Tauri
stars. This hypothesis is supported by the asymmetric line 
profile of the [OI] lines, which display an extended blue wing but
miss the red wing.
These results, together with the fact that the forbidden lines are
known to form at different 
distances from the star, strongly argue in favor of a disk seen at a
geometry markedly different from edge-on.
As in the case of Par-Lup3-4, the 10\% width of the \Ha\ emission,
with values in the range from 265 to 
300~km$\cdot$s$^{-1}$, indicates that \Ha\ is dominated by the
accretion component formed near the surface of the star.



 Barrado y Navascu\'es et al. (\cite{Barrado04.1}) favour the
hypothesis of an edge-on disk, as an explanation for the puzzling
aspects of LS-RCrA~1, namely the lack of near-infrared (NIR) excess
combined with accretion, the unusually prominent outflow signatures
without high-velocity components or asymmetries, the very broad \Ha, and
the sub-luminosity. Nevertheless, we think that there are also
explanations for all these features in the framework of the non edge-on
disk 
hypothesis. No NIR excesses are expected for very low mass stars and
brown dwarfs, as has been modeled by Natta \& Testi (\cite{Natta01.1}) and
has been confirmed by Barrado y Navascu\'es et al. (\cite{Barrado04.3}),
except for very few objects;
such an excess is predicted for wavelengths longer than $\sim$3$\mu$m.
Our high resolution spectra show asymmetries on the forbidden line
profiles. The very broad \Ha\ profile indicates an unimpeded view to the
close proximity of the stellar surface, where the largest velocities of
the \Ha-emitting gas arise (Appenzeller et
al. \cite{Appenzeller05.1}). Concerning the sub-luminosity, we
still support the hypothesis suggested by Fern\'andez \& Comer\'on
(\cite{Fernandez01.1}) that, at these low masses, strong mass accretion
might have an important effect on the position of the star on the HR
diagram.

 The central velocities measured for the forbidden lines differ
notably. The lower velocities are found for the [OI] lines, while the
[NII] lines give the highest values (see
Table~\ref{tab:veloc_Par_LS}). The [OI] $\lambda$ 6300 line has a
critical density higher than that of [NII] 6583, and roughly 100 times
that of the [SII] lines (Hartigan et al. \cite{Hartigan95.1}).  Hartigan
et al. analyzed a sub-sample of four stars for which they detect
emission from three lines, [OI] $\lambda$ 5577, [OI] $\lambda$ 6300, and
[SII] $\lambda$ 6731, that have very different critical
densities. The low velocity component of the three emission lines show a
correlation in which the lower velocities correspond to the line with
highest critical density ([OI]\,$\lambda$5577); a similar behaviour is
observed for 
LS-RCrA~1. They found this correlation to be consistent with
acceleration in a disk wind, but also with an origin in an accretion
column, because in either case the lines with higher critical density
(like [OI]) form closer to the disk than the lines with low critical
density, and the flow accelerates as it rises from the disk. 
Hirth et
al. (\cite{Hirth97.1}) found that for a sample of 12 T~Tauri stars,
located at $\sim$120-140~pc, the
centroid of the [OI] 6300 emission is located at an average distance of
0\farcs2 from the star, whereas that of the [SII]\,$\lambda$6731 and
[NII]\,$\lambda$6583 lines are factors of 3 and 3.5 times further away,
respectively.


 The bumps observed at about 50\kms\ on the [SII] and [NII] lines may be
due to either matter ejected closer to our line of sight or to a faster
knot. Similar bumps have been observed, e.g., for the T Tauri stars
DK~Tau, GG~Tau ad IP~Tau (Hartigan et al. \cite{Hartigan95.1}).


%

\section{Conclusions}

  We report the discovery of a jet emanating from the very low mass star
Par-Lup3-4 (M5), and we confirm previous evidence of strong mass loss
from another very low mass star, LS-RCrA~1 (M6.5 or later), most
probably in the form of a jet or disk wind.

 The line ratios of the forbidden lines of the jet of Par-Lup3-4 point
to a low excitation jet. The double-peaked [SII] emission,
centered on the LSR velocity of the Lupus~3 cloud in this region, allows
us to set lower limits for the jet inclination: angles below 8\degr, 
with respect to the plane of the sky,
would imply unlikely velocities above 150 km$\cdot$s$^{-1}$.
With such inclination only a very flared disk would hide the star. The
large 10\% width is
attributed to the accretion related regions, which lie very close to or
on the stellar surface, suggesting that the large \Ha\ equivalent width
measured for this object is not due to the selective
blocking of the central object by an edge-on disk.

 \Ha\/ emission, coming for an object located at 4\farcs2 from
Par-Lup3-4, has been detected in the two spectra taken at PA $\sim$
33$^{\circ}$. The line profile, different from that of Par-Lup3-4,
resembles that of other pre-main sequence objects. Upper limits for its
brightness at visible and near infrared wavelengths suggest that, if
associated with the Lupus~3 star forming region, it could be a young, very
low mass brown dwarf.

 All the forbidden lines that we have measured for LS-RCrA~1 are
blueshifted with respect to the LSR velocity of the star. The emission from
the receding part of the jet seems to be hidden by a non edge-on disk; a
hypothesis that is supported by the fact that we detect \Ha\ emission
coming from the accretion related regions located close to the
surface of the star.
The velocities of the [OI] and [SII] forbidden emission lines are
ordered inversely with their respective critical densities. This has
been interpreted by Hartigan et al. (\cite{Hartigan95.1}), for more
massive classical T~Tauri stars, as acceleration from the most dense
regions, close to the star, what would be consistent with acceleration
in a disk wind, but also with an origin in an accretion column.

 If both Par-Lup3-4 and LS-RCrA~1 have no edge-on disk, an alternative
explanation is required in order to explain their unusual low
luminosities. Strong accretion has been suggested to modify the position
of classical T Tauri stars on the HR diagram (Hartmann et
al. \cite{Hartmann97.1}, Siess et al. \cite{Siess97.1}). 
Extending the modelling of the accretion effects towards the
lowest stellar and substellar masses may indicate whether or not this is
a viable explanation for the observed properties of these objects.


\begin{acknowledgements}

  We acknowledge the ESO staff who carried out the service mode
observations, the ESO User Support Group for their valuable assistance
in the preparation of our observations, and the ESO Data Flow Operations
Group for the preparation of our data package. A. Kaufer, S. D'Odorico
and L. Kaper, authors of the \uves\ User Manual, are also acknowledged,
as well as those who prepared the \uves\ web pages.  Fruitful
discussions with Eike Guenther, 
Jochen Eisl\"offel, Jens Woitas, Enrique P\'erez, Reinhard Mundt
and Ferdinando Patat were very helpful, as were 
comments from David Barrado y Navascu\'es and from David Butler. MF
acknowledges ESO and the Th\"uringer
Landessternwarte (Germany) for their hospitality. She received support
from the Deutsches Zentrum f\"ur Luft- und Raumfahrt (DLR),
F\"orderkennzeichen 50 OR 0401, and from the Spanish grant
AYA2004-05395.  This work has made use of the Digitized Sky Surveys,
produced at the Space Telescope Science Institute under U.S. Government
grant NAG W-2166, and of the NASA/IPAC Infrared Science Archive, which
is operated by the Jet Propulsion Laboratory, California Institute of
Technology, under contract with the National Aeronautics and Space
Administration.

\end{acknowledgements}

\end{document}